\documentclass[aps,prl,twocolumn,amsfonts,showpacs,superscriptaddress]{revtex4} 
\usepackage{bm}
\usepackage{epsfig,amsopn}
\usepackage{graphicx,color}
\usepackage{amsmath,amssymb, amsfonts, amsthm}
\usepackage{braket}
\newcommand{\sectionprl}[1]{{\par\it #1.---}}

\newcommand{\Tr}[0]{ \textrm{Tr}}

\newtheorem{lemma}{Lemma}

\newcommand{\todayd}{\the\year/\the\month/\the\day}

\newcommand{\bel}{\begin{easylist}}
\newcommand{\eel}{\end{easylist}}

\def \({\left(}
\def \){\right)}
\def \[{\left[}
\def \]{\right]}

\newcommand{\sumtwo}[2]%
{\mathop{\sum_{#1}}_{#2}}
\newcommand{\sumthree}[3]%
{\mathop{\mathop{\sum_{#1}}_{#2}}_{#3}}
\newcommand{\sumfour}[4]%
{\mathop{\mathop{\mathop{\sum_{#1}}_{#2}}_{#3}}_{#4}} 
\newcommand{\prodtwo}[2]%
{\mathop{\prod_{#1}}_{#2}}
\newcommand{\mintwo}[2]%
{\mathop{\min_{#1}}_{#2}}
\newcommand{\maxtwo}[2]%
{\mathop{\max_{#1}}_{#2}}
\newcommand{\maxthree}[3]%
{\mathop{\mathop{\max_{#1}}_{#2}}_{#3}}
\newcommand{\limtwo}[2]%
{\mathop{\lim_{#1}}_{#2}}
\newcommand{\suptwo}[2]%
{\mathop{\sup_{#1}}_{#2}}
\newcommand{\supthree}[3]%
{\mathop{\mathop{\sup_{#1}}_{#2}}_{#3}}
\newcommand{\supfour}[4]%
{\mathop{\mathop{\mathop{\sup_{#1}}_{#2}}_{#3}}_{#4}} 
\newcommand{\inftwo}[2]%
{\mathop{\inf_{#1}}_{#2}}
\newcommand{\infthree}[3]%
{\mathop{\mathop{\inf_{#1}}_{#2}}_{#3}}
\newcommand{\inffour}[4]%
{\mathop{\mathop{\mathop{\inf_{#1}}_{#2}}_{#3}}_{#4}} 

\newcommand\calE{{\cal E}}
\newcommand\calF{{\cal F}}

\newcommand\calH{{\cal H}}
\newcommand\calI{{\cal I}}

\newcommand\calK{{\cal K}}
\newcommand\calL{{\cal L}}

\newcommand\calR{{\cal R}}

\newenvironment{proofof}[1]{\vspace*{5mm} \par \noindent
{\bf Proof of #1:\hspace{2mm}}}{\endproof
}

\begin{document}
\title{Coherence-variance uncertainty relation and coherence cost for quantum measurement under conservation laws}
\author{Hiroyasu Tajima}
\affiliation{Yukawa institute for theoretical physics, Kyoto University
Oibuncho Kitashirakawa Sakyo-ku Kyoto city, Kyoto, 606-8502, Japan }

\author{Hiroshi Nagaoka}
\affiliation{University of Electro-Communications,
1-5-1 Chofugaoka, Chofu, Tokyo, 182-8585, Japan}

\begin{abstract}
Uncertainty relations are one of the fundamental principles in physics.
It began as a fundamental limitation in quantum mechanics, and today the word {\it uncertainty relation} is a generic term for various trade-off relations in nature.
In this letter, we improve the Kennard-Robertson uncertainty relation, and clarify how much coherence we need to implement quantum measurement under conservation laws.
Our approach systematically improves and reproduces the previous various refinements of the Kennard-Robertson inequality.
As a direct consequence of our inequalities, we improve a well-known limitation of quantum measurements, the Wigner-Araki-Yanase-Ozawa theorem. This improvement gives an asymptotic equality for the necessary and sufficient amount of coherence to implement a quantum measurement with the desired accuracy under conservation laws.
\end{abstract}

\pacs{
03.65.Ta, 
03.67.-a, 
05.30.-d, 
42.50.Dv, 
}

\maketitle

\sectionprl{Introduction}
Uncertainty relations are one of the most fundamental limitations in physics.
Historically, it starts with Heisenberg's famous comment \cite{Heisenberg} and Kennard's proof \cite{Kennard}.
It was immediately generalized to the following well-known inequality by Robertson \cite{Robertson}:
\begin{align}
V_{\rho}(A)V_{\rho}(B)\ge\frac{|\left<[A,B]\right>_{\rho}|^2}{4}
\end{align}
After Robertson's discovery, similar trade-off relations have been vigorously explored by many physicists \cite{Luo-unc,Yanagi2010,Yanagi2011,Gibilisco2011,Marvian-thesis,Frowis2015,
Mandelstam1945,Margolus1998,Pires2016,Shiraishi2017,Ito2018,Funo2019,Ito2018-2,
Ozawa2003(measurement),Watanabe2011,Fujikawa2012,Korezekwa2014,
Fuchs1996,Fuchs1998,Miyadera2006,Maccone2006,Buscemi2008,ozawa1,ozawa2,Karasawa2007,Karasawa2009,Tajima2018,Tajima2019,Barato2015, Gingrich2016, Pigolotti2017}.
These quests resulted in various refinements of the Robertson inequality \cite{Luo-unc,Yanagi2010,Yanagi2011,Gibilisco2011,Marvian-thesis,Frowis2015} and various different types of trade-offs relations \cite{Mandelstam1945,Margolus1998,Pires2016,Shiraishi2017,Ito2018,Funo2019,Ito2018-2,
Ozawa2003(measurement),Watanabe2011,Fujikawa2012,Korezekwa2014,
Fuchs1996,Fuchs1998,Miyadera2006,Maccone2006,Buscemi2008,ozawa1,ozawa2,Karasawa2007,Karasawa2009,Tajima2018,Tajima2019,Barato2015, Gingrich2016, Pigolotti2017}; the time-energy uncertainty (speed-limit) in the state control of quantum and classical systems \cite{Mandelstam1945,Margolus1998,Pires2016,Shiraishi2017,Ito2018,Funo2019,Ito2018-2}; the error-disturbance trade-off relations in quantum measurements \cite{Ozawa2003(measurement),Watanabe2011,Fujikawa2012,Korezekwa2014}; the information gain and disturbance trade-off in quantum measurements \cite{Fuchs1996,Fuchs1998,Miyadera2006,Maccone2006,Buscemi2008}; 
the error and the fluctuation trade-off relations in implementing unitary operation on quantum systems \cite{ozawa1,ozawa2,Karasawa2007,Karasawa2009,Tajima2018,Tajima2019}; and the thermodynamic uncertainty that is a kind of dissipation and fluctuation trade-off relation in stochastic thermodynamics \cite{Barato2015, Gingrich2016, Pigolotti2017}.
Today, {\it uncertainty relation} is a term that represents a very broad field consisting of these various type inequalities.

One of the most famous and important applications of the Robertson inequality is the Wigner-Araki-Yanase-Ozawa (WAY-Ozawa) theorem \cite{OzawaWAY}.
This is a quantitative refinement of the famous Wigner-Araki-Yanase (WAY) theorem \cite{Wigner1952,Araki-Yanase1960,Yanase1961}, a limitation that always holds when we perform quantum measurements under the conservation law.
Roughly speaking, the WAY-Ozawa theorem asserts that:
{\it In order to precisely measure the instantaneous value of a physical quantity which does not commutes with the conserved quantity, we have to prepare a considerable amount of the fluctuation of the conserved quantity.}

Since the WAY-Ozawa theorem is a direct consequence of the Kennard-Robertson inequality, it is natural to try to improve the theorem using the refiments of the Kennard-Robertson inequality.
In fact, several refinements of Robertson inequality \cite{Luo-unc,Marvian-thesis} have been used to improve the WAY-Ozawa theorem, and several lower bounds on the amount of quantum coherence required for quantum measurements under the conservation laws were given \cite{Korezekwa-thesis}.
In spite of the progress, we do not still reach an exact understanding on the amount of quantum coherence required to implement quantum measurements under conservation laws.
It is not known how tight these lower bounds are, and any upper bound for the sufficient amount of coherence has not been given so far.

In this letter, we revisit and refine the Kennard-Robertson uncertainty relation based on the information geometry, and give a solution for the above problem.
In solving the problem, we evaluate the amount of the required coherence for measurement in terms of an {\it equality}.
That is, we give an asymptotic equality for the quantum coherence that is necessary and sufficient to implement the measurement with an error $\epsilon$.
This asymptotic equality shows that the necessary and sufficient amount of coherence is asymptotically written by only two amounts. The first one is the error $\epsilon$, and the second one is the norm of the commutator between the conserved quantity $A$ and the measured quantity $B$.

For the above goal, we investigate two systematic approaches.
First, we systematically improve the Robertson inequality with quantum information geometric method.
Our approach systematically improves and reproduces the previous various refinements of the Kennard-Robertson uncertainty relation \cite{Luo-unc,Yanagi2010,Yanagi2011,Gibilisco2011,Marvian-thesis,Frowis2015}.
As a direct consequence of this systematic approach, we improve the WAY-Ozawa theorem. 
Our improved WAY-Ozawa theorem gives a universal lower bound for the amount of coherence required to implement quantum measurement under conservation laws. The bound is strict tighter than the previous refinement of the WAY-Ozawa inequality \cite{Korezekwa-thesis}. 

Next, we show the optimality of our improved WAY-Ozawa theorem.
In order to show the optimality, we investigate another systematically method which constructs an indirect measurement process which realizes the desired measurement within the desired accuracy with the minimal sufficient quantum coherence.
The construction gives upper bounds for sufficient quantum coherence to implement measurement under conservation laws.
The upper and lower bounds always match asymptotically in the region where the implementation error is small.   
Combining the upper bounds and the improved WAY-Ozawa theorems, we obtain asymptotic equalities for coherence costs of quantum measurement under conservation laws with various measures of coherence.
The asymptotic equalities quantitatively show a simple relation among measurement theory, conservation laws and quantum coherence.

\begin{figure}
\includegraphics[width=9cm]{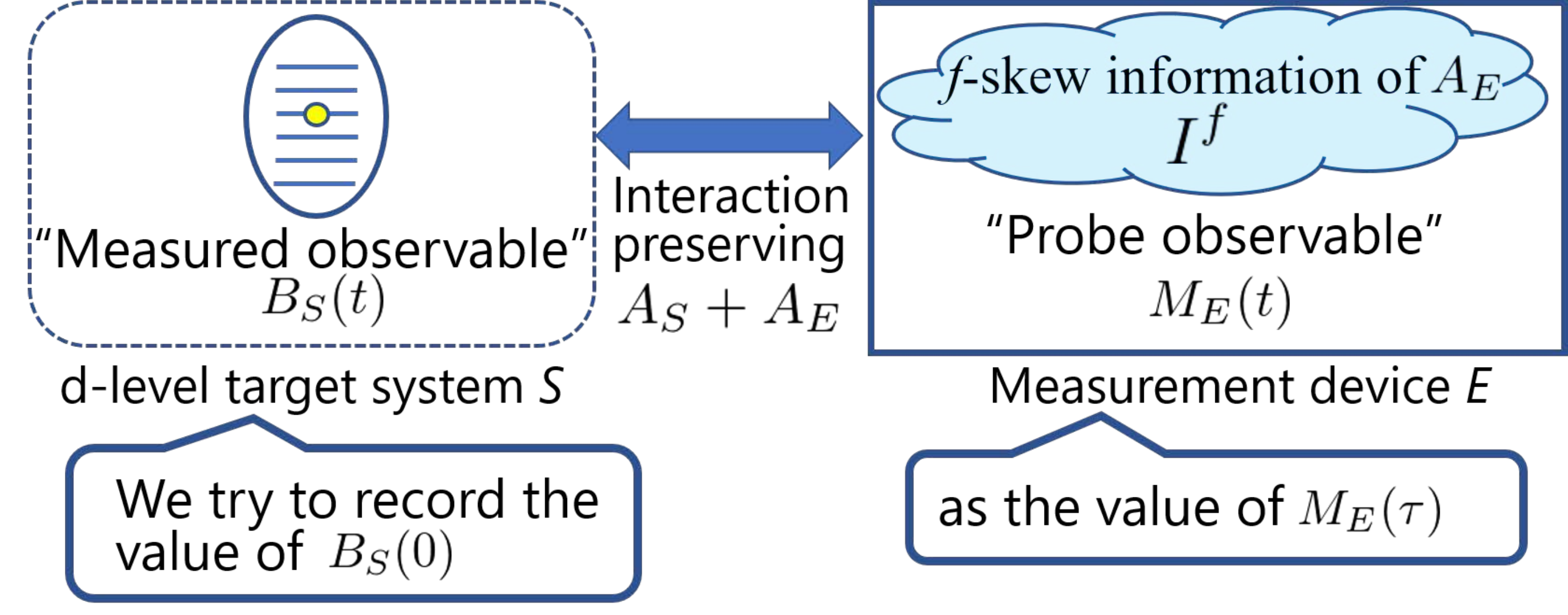}
\caption{Schematics of the setup of the improved WAY-Ozawa theorem.
We follow the original setup of the WAY-Ozawa theorem.
We perform a measurement on $S$ as an indirect measurement process with another quantum system $E$ from $t=0$ to $t=\tau$.
Conserved quantities exist in the system, of which one of them is denoted by $A_S$. The external system is supposed to exhibit the same type of quantity $A_E$, and the operator $A_S + A_E$ is a conserved quantity for the time-evolution of the entire composite system.
We aim to measure the spontaneous value of $B_S$ that does not commute with the quantity $A_S$.
We take the {\it probe observable} $M_E$ that commutes with $A_E$.
We consider the Heisenberg representation of $B_S$ and $M_E$, and define $M_E(\tau)$ as the {\it measured value} of $B_S(0)$.
Then, we consider the error of the measurement as the expectation value of $(B_S(0)-M_E(\tau))^2$.
We consider the necessary condition of the amount of the $f$-skew information to make the error small.}\label{FigO-WAY}
\label{setup}
\end{figure}

\sectionprl{Quantum measurement under conservation law}
We firstly introduce the set up of the implementation of quantum measurement under the conservation law of some physical quanitity $A$ (Fig. \ref{FigO-WAY}). 
The quantity $A$ is energy, for example.
We follow the traditional set up that is used in the original paper of WAY-Ozawa theorem \cite{OzawaWAY}.

Let us take a quantum system $S$ as the target system whose Hilbert space is ${\cal H}_S$.
We refer to the operator of the conserved quantity $A$ on $S$ as $A_S$.
Without loss of generality, we can assume that the minimum eigenvalue of $A_S$ is zero.
We try to measure the value of an Hermitian operator $B_S$ on $S$ as the following indirect measurement process that starts at $t=0$ and finishes at $t=\tau$:
\begin{itemize}
\item[Step 1]{ We take an external quantum system $E$ whose Hilbert space and initial state are $\calH_E$ and $\rho_E$, respectively. We refer to the operator of $A$ on $E$ as $A_E$. We also define the {probe observable} $M_E$ on $E$ satisfying $[M_E,A_E]=0$.}
\item[Step 2]{We take an Hermitian operator $H_{SE}$ on $SE$, and perform an unitary dynamics $U_{SE}:=e^{-iH_{SE}\tau}$ on $SE$. We assume that $U_{SE}$ satisfies the conservation law of $A$, i.e., $[U_{SE},A_S+A_E]=0$.}
\end{itemize}
Because the set $(\calH_E,A_E,M_E,\rho_E,U_{SE})$ completely determines the above indirect measurement process, we refer to the set as the implementation set $\calI$.
In order to define the error of the measurement, we use the Heisenberg picture.
In the Heisenberg picture with the original state $\rho_S\otimes\rho_E$ at $t=0$, we shall write $B_S(0)=B_S\otimes1_E$, $M_E(0)=1_S\otimes M_E$, $B_S(\tau)=U^{\dagger}_{SE}(B_S\otimes1_E)U_{SE}$ and $M_E(\tau)=U^{\dagger}_{SE}(1_S\otimes M_E)U_{SE}$.
Then we determine the {\it noise operator} $N_{SE}$ as the difference the mesured observable at $t=0$ and the the probe observable at $t=\tau$, and define the error of the measurement $\epsilon_{\calI}(\rho_S)$ as the square root of the expectation value of $N_{SE}^2$:
\begin{align}
\epsilon_{{\cal I}}(\rho_S)^2&:=\sqrt{\Tr[(\rho_S\otimes\rho_E)N_{SE}^2]},\\
N_{SE}&:=B_S(0)-M_{E}(\tau).
\end{align}
As we discuss in the supplementary materials, our results are also valid in more general framework described by $G$-covariant operations used in resource theory of asymmetry \cite{Bartlett, Korezekwa-thesis, Gour, Marvian, Marvian2016, Marvian-thesis,Takagi2018, Marvian2018, Lostaglio2018,Takagibunken1}.

\sectionprl{Measures of coherence: metric adjusted skew informations}
Next, we introduce the measures of coherence used in this letter.
We employ Luo's definition \cite{Luo2005}, which is one of the most common approaches.
In Luo's approach, we treat the part of the variance due to the quantum superposition as a measure of the amount of coherence.
Let us formally divide the variance $V_{\rho}(A):=\Tr[\rho A^2]-\Tr[\rho A]^2$ of a Hermitian operator $A$ as follows:
\begin{align}
V_{\rho}(A)=Q_{\rho}(A)+C_{\rho}(A)
\end{align}
Then, if the quantities $Q_{\rho}(A)$ and $C_{\rho}(A)$ express the quantum part and classical part of the variance, respectively, the quantity $Q_{\rho}(A)$ satisfies the following properties:
\begin{itemize}
\item[P1]{The quantity $Q_{\rho}(A)$ is nonnegative, and equal to or smaller than $V_{\rho}(A)$. (i.e., $0\le C_{\rho}(A)\le V_{\rho}(A)$)}
\item[P2]{If the fluctuation of $A$ in $\rho$ is purely due to classical mixture, namely if $\rho$ commutes with $A$, then $Q_{\rho}(A)$ is equal to $0$.}
\item[P3]{If the fluctuation of $A$ in $\rho$ is purely due to quantum superposition, namely if $\rho$ is pure, then $Q_{\rho}(A)$ is equal to $V_{\rho}(A)$.}
\item[P4]{The quantity $Q_{\rho}(A)$ is decreased by classical mixture. Namely, $Q_{\rho}(A)$ satisfies $Q_{\sum_ip_i\rho_i}(A)\le\sum_ip_iQ_{\rho_i}(A)$ for any probability $\{p_i\}$ and density matrices $\{\rho_i\}$.}
\end{itemize}
There are so many (actually infinite) quantities $Q_{\rho}(A)$ satisfy the properties P1--P4.
It is well known that most of them are described as the following quantity given by Hansen, named {\it metric adjusted skew information} \cite{Hansen2008}:
\begin{align}
I^{f}_{\rho}(A):=\frac{f(0)}{2}\sum_{i,j}\frac{(p_i-p_j)^2}{p_jf(p_i/p_j)}|\bra{i}A\ket{j}|^2.
\end{align}
Here,  $\{p_i,\ket{i}\}$ are the eigenvalues and eigenvectors of $\rho$, and $f$ is a standard operator monotonic function satisfying the following properties:
\begin{itemize}
\item[Q1]{For any Hermitian operators $A$ and $B$, the inequality $0\le A\le B$ implies $f(A)\le f(B)$.}
\item[Q2]{$f(x)=xf(1/x)$.}
\item[Q3]{$f(1)=1$.}
\end{itemize}
We refer to the quantity $I^{f}_{\rho}(A)$ with given $f$ as $f$-skew information, and refer to the family of the $f$-skew information as the metric adjusted skew information.

For arbitrary $f$ satisfying Q1--Q3, the $f$-skew information satisfies P1--P4 \cite{Hansen2008}.
Also, each $f$-skew information is measurable in experiments \cite{Shitara2016}. 
Therefore, we can interpret each $I^{f}_{\rho}(A)$ as a measure of coherence.
The traditional coherence measures can be reproduced by putting a function of a specific form in $f$.
For example, Wigner-Yanase skew information \cite{Wigner-Yanase} and SLD skew information are expressed by the following functions:
\begin{align}
f_{WY}(x):=\left(\frac{\sqrt{x}+1}{2}\right)^2,\enskip
f_{SLD}(x):=\frac{1+x}{2}.
\end{align}
Because of these reasons, recently the $f$-skew information is used as a common series of measures of quantum coherence.
The $f$-skew information also plays an important role in resource theory of asymmetry \cite{Bartlett, Korezekwa-thesis, Gour, Marvian, Marvian2016, Marvian-thesis,Takagi2018, Marvian2018, Lostaglio2018,Takagibunken1}. For example, it is shown that each $f$-skew information is resource measure of asymmetry \cite{Takagibunken1}.

It is noteworthy that the metric adjusted skew informations are closely related to the quantum Fisher informations. 
Let us consider the family of the quantum states $\{\rho_t\}$ that is parametrized by a single real number $t$.
Then, the quantum Fisher informations are described as follows:
\begin{align}
\calF^{f}(\rho_t):=\left<L,L\right>^{f}_{\rho},
\end{align}
where $\left<...,...\right>^{f}_{\rho}$ is the inner product:
\begin{align}
\left<A,B\right>^{f}_{\rho}:=\Tr[A^{\dagger}m_{f}(\calL_{\rho},\calR_{\rho})(B)],
\end{align}
and $L$ is the operator satisfying 
\begin{align}
\frac{\partial \rho_t}{\partial t}=m_{f}(\calL_\rho,\calR_\rho)(L)
\end{align}
where $m_f(x,y)=yf(x/y)$, $\calL_{\rho}$ and $\calR_{\rho}$ are the superoperators multiplying $\rho$ from left and right, i.e., $\calL_{\rho}X=\rho X$ and $\calR_{\rho}X=X\rho$.
In general, $f$-skew information is described with $\calF^{f}(\rho_t)$ in case of $\rho_{t}:=e^{-iAt}\rho e^{iAt}$ as follows \cite{Hansen2008}:
\begin{align}
I^{f}_{\rho}(A)=\frac{f(0)}{2}\calF^{f}(\rho_{t})|_{t=0}.\label{skew-Fisher}
\end{align}

\sectionprl{Improved WAY-Ozawa theorem}
Now, we have prepared to introduce our first main result, i.e., the improved versions of WAY-Ozawa theorem.
The following inequality always holds for arbitrary $\calI$, $B_S$ and $\rho_S$:
\begin{align}
\epsilon_{\calI}(\rho_S)^2\ge\frac{|\left<[A_S,B_S]\right>_{\rho_S}|^2}{4(I^{f_{SLD}}_{\rho_S}(A_S)+I^{f_{SLD}}_{\rho_E}(A_E))}\label{skew-WAY}
\end{align}
We can see the original WAY-Ozawa theorem by substituting $V_{\rho_S}(A_S)+V_{\rho_E}(A_E)$ for $I^{f_{SLD}}_{\rho_S}(A_S)+I^{f_{SLD}}_{\rho_E}(A_E)$ the in the above inequality.
Therefore, due to the property P1, our improved WAY-Ozawa theorem gives always tigher bound than the original bound.
It is also northworthy that the inequality \eqref{skew-WAY} is always tighter than Korzekwa's refinement of the WAY-Ozawa theorem in Ref. \cite{Korezekwa-thesis}.
We can see Korzekwa's refinement by substituting $2(I^{f_{WY}}_{\rho_S}(A_S)+I^{f_{WY}}_{\rho_E}(A_E))$ for $I^{f_{SLD}}_{\rho_S}(A_S)+I^{f_{SLD}}_{\rho_E}(A_E)$ in \eqref{skew-WAY}.
Due to the inequality $I^{f_{SLD}}_{\rho}(A)\le 2I^{f_{WY}}_{\rho}(A)$ holds for arbitrary $\rho$ and $A$ \cite{Petz2010}, \eqref{skew-WAY} is always tighter than orzekwa's refinement.

\sectionprl{Coherence cost for quantum measurement}
Let us introduce our second main result, i.e., the asymptotic equality of coherence cost for quantum measurement.
We define the worst error in the measurement of $B_S$ implemented by the implementation set $\calI$ as the maximum of the error $\epsilon_{{\cal I}}(\rho_S)$ among all initial states $\rho_S$:
\begin{align}
\epsilon_{{\cal I}}:=\max_{\rho_S}\epsilon_{\calI}(\rho_S)\label{maxerror}
\end{align}
With using the maximal error, we define the minimal sufficient amount of $f$-skew information to measure
$B_S$ within the error $\epsilon$ as follows:
\begin{align}
I^{f}_{\epsilon,\mbox{cost}}:=\min_{\calI:\epsilon_{\calI}\le\epsilon}I^{f}_{\rho_E}(A_E).\label{costformeas}
\end{align}
Then, the following inequality holds for arbitrary $f$ satisfying Q1--Q3:
\begin{align}
\frac{\|[A_S,B_S]\|}{\epsilon\sqrt{2/f(0)}}-\frac{\|A_S\|}{2}
\le\sqrt{I^{f}_{\epsilon,\mbox{cost}}}\le\frac{\|[A_S,B_S]\|}{2\epsilon}+\|A_S\|\label{UandL}
\end{align}
where the upper bound holds for small $\epsilon$ satisfying $0<\epsilon\le\frac{\|[A_S,B_S]\|}{8\|A_S\|}$.
We prove the lower and upper bounds in \eqref{UandL} in the derivation section and the supplementary materials, respectively.
From \eqref{UandL}, we obtain the following asymptotic equality for arbitrary $f$ satisfying $f(0)=1/2$ and Q1--Q3:
\begin{align}
\sqrt{I^{f}_{\epsilon,\mbox{cost}}}=\frac{\|[A_S,B_S]\|}{2\epsilon}+O(\|A_S\|),\enskip \epsilon\rightarrow0.\label{skew-as}
\end{align}

\sectionprl{Coherence-variance uncertainty relations}
To derive the inequalities \eqref{skew-WAY} and \eqref{UandL}, we use the following lemma
\begin{lemma}
Let us take physical quantities $A$ and $B$ as Hermitian operators. We also take an arbitrary state $\rho$.
Then, the following trade-off relation between $f$-skew information of $A$ and $f$-variance of $B$ for arbitrary $f$: 
\begin{align}
I^{f}_{\rho}(A)V^{f}_{\rho}(B)\ge\frac{f(0)}{2}|\left<[A,B]\right>_{\rho}|^2\label{skew-variance}
\end{align}
where the $f$-variance $V^{f}$ is defined as follows:
\begin{align}
V^{f}_{\rho}(A):=\left<A_0,A_0\right>^{f}_{\rho}
\end{align}
where $A_0:=A-\left<A\right>_\rho$.
\end{lemma}
Due to the property P1 and $V_{\rho}(A)\ge V^{f}_{\rho}(A)$ \cite{notev}, the above inequality is {\it always} tighter than the original Robertson inequality whenever $f(0)=1/2$ holds.
Moreover, as we show in the supplemetary materials, we can improve and reproduce the previous refinements of Robertson's uncertainty relations in Refs.\cite{Luo-unc,Yanagi2010,Yanagi2011,Gibilisco2011,Marvian-thesis,Frowis2015} from \eqref{skew-variance}.

The inequality \eqref{skew-variance} is given as follows.
For an arbitrary family of the state $\{\rho_t\}$, we obtain 
\begin{align}
\left|\frac{\partial\left<B\right>_{\rho_t}}{\partial t}\right|^2&=\left|\left<B_0,L\right>^{f}_{\rho_t}\right|^2\nonumber\\
&\le\left<B_0,B_0\right>^{f}_{\rho_t}\left<L,L\right>^{f}_{\rho_t}\nonumber\\
&=V^{f}_{\rho_t}(B)\calF^{f}(\rho_t).\label{Fisher-variance}
\end{align}
Here, we use the definition of $L$ in the first line, and use the Cauchy-Schwartz inequality in the second line.
Substituting $e^{-iAt}\rho e^{iAt}$ for $\rho_t$ in \eqref{Fisher-variance},  and, we obtain 
\begin{align}
\calF^{f}_{\rho}(A)V^{f}_{\rho}(B)\ge|\left<[A,B]\right>_{\rho}|^2\label{Fisher-variance-uncer}.
\end{align}
Substituting \eqref{skew-Fisher} in \eqref{Fisher-variance-uncer}, we obtain \eqref{skew-variance}.

\sectionprl{Derivation}
Finally, we derive our main results \eqref{skew-WAY} and \eqref{UandL}.
The inequality \eqref{skew-WAY} is a direct consequence of the coherence-variance uncertainty relation \eqref{skew-variance}.
The derivation is the same as the derivation of the original WAY-Ozawa theorem from the Robertson inequality.
From \eqref{skew-variance} and the fact $\epsilon_{\calI}(\rho_S)^2$ is larger than $V_{\rho_S\otimes\rho_E}(A_S\otimes A_E)$, we obtain
\begin{align}
\epsilon_{\calI}(\rho_S)^2&\ge
\frac{f(0)|\left<[N_{SE},A_S+A_E]\right>_{\rho_S}|^2}{2(I^{f}_{\rho_S}(A_S)+I^{f}_{\rho_E}(A_E))}.\label{skew-WAYpre}
\end{align}
Then, we obtain \eqref{skew-WAY} from \eqref{skew-WAYpre} with using the following transformation:
\begin{align}
&\left<[N_{SE},A_S+A_E]\right>_{\rho_S\otimes\rho_E}\nonumber\\
&=\left<[(B_S\otimes1_E)-U^{\dagger}_{SE}(1_S\otimes M_E)U_{SE},A_S+A_E]\right>_{\rho_S\otimes\rho_E}\nonumber\\
&=\left<[(B_S\otimes1_E),A_S+A_E]\right>_{\rho_S\otimes\rho_E}\nonumber\\
&\enskip\enskip-\left<[1_S\otimes M_E,A_S+A_E]\right>_{U_{SE}(\rho_S\otimes\rho_E)U^{\dagger}_{SE}}\nonumber\\
&=\left<[B_S,A_S]\right>_{\rho_S}.\label{ozawa-trans}
\end{align}
Here we use $[U_{SE},A_S+A_E]=0$ and $[M_E,A_E]=0$ in the second line.
Substituting \eqref{ozawa-trans} and $f(x)=f_{SLD}(x)$ in \eqref{skew-WAYpre}, we obtain \eqref{skew-WAY}.

Next, we derive the lower bound in \eqref{UandL}.
Subsituting \eqref{ozawa-trans} in \eqref{skew-WAYpre} and using the property P1 and $V_{\rho_S}(A_S)\le\|A_S\|^2/4$, we obtain
\begin{align}
\epsilon_{\calI}(\rho_S)^2\ge\frac{f(0)|\left<[A_S,B_S]\right>_{\rho_S}|^2}{2(\frac{1}{4}\|A_S\|^2+I^{f}_{\rho_E}(A_E))}
\end{align}
By maximizing the both side through all of $\rho_S$, we obtain
\begin{align}
\epsilon_{\calI}^2\ge\frac{f(0)\|[A_S,B_S]\|^2}{2(\frac{1}{4}\|A_S\|^2+I^{f}_{\rho_E}(A_E))}
\end{align}
Therefore, we obtain 
\begin{align}
\sqrt{\frac{f(0)}{2}}\frac{\|[A_S,B_S]\|}{\epsilon_{\calI}}&\le\sqrt{I^{f}_{\rho_E}(A_E)+\frac{\|A_S\|^2}{4}}\nonumber\\
&\le\sqrt{I^{f}_{\rho_E}(A_E)}+\frac{\|A_S\|}{2}.
\end{align}
Hence, we obtain the following lower bound:
\begin{align}
\sqrt{I^{f}_{\epsilon,\mbox{cost}}}\ge\sqrt{\frac{f(0)}{2}}\frac{\|[A_S,B_S]\|}{\epsilon}-\frac{1}{2}\|A_S\|\label{lowercost}
\end{align}

\sectionprl{Conclusion}
In this letter, we systematically improve the Kennard-Robertson uncertainty from a quantum information geometric method.
Our method give the trade-off inequalities that restrict the product of $f$-skew information of physical quantity $A$ and the dispersion of physical quantity $B$ by the commutator between $A$ and $B$. 
Because each $f$-skew information is a coherence measure, the obtained inequalities represent trade-off relations between the quantum coherence of $A$ and the fluctuation of $B$.
These inequalities are tighter than the original Kennard-Robertson inequality for arbitrary $f$ satisfying $f(0)=1/2$.
The inequalities give, as a direct consequence, the lower bounds on the amount of coherence necessary to make measurements under the conservation law.
These lower bounds are tighter than the original WAY-Ozawa theorem for arbitary $f$ satisfying $f(0)=1/2$ and the previous refinement WAY-Ozawa theorem.
We also obtain the upper bound of the amount of coherence sufficient to make measurements under the conservation law.
These bounds correspond to the case of $f(0)=1/2$, giving an asymptotic equation.
The asymptotic equalities for the amount of necessary and sufficient coherence to implement a quantum measurement with the desired accuracy under conservation laws for $f$-skew informations satisfying $f(0)=1/2$.
The asymptotic equalities quantitatively reveal a simple link among measurement theory, conservation laws and quantum coherence.

Finally, we point out the similarity between the asymptotic equalities in this letter and the asymptotic equality in Ref.\cite{Tajima2019}.
The asymptotic equality in Ref.\cite{Tajima2019} clarifies the coherence cost to implement an arbitrary unitary operation within desired error under conservation laws.
The asymptotic equality has exactly the same form as \eqref{skew-as} in this letter; the coherence cost for unitary gates is asymptotically inversely proportional to the error.
Therefore, the coherence costs for quantum measurement and unitary gates have the same structure under the restriction of conservation laws.
The fact that the same asymptotic equations hold for two seemingly different objects, quantum measurements and unitary gates, is truly surprising.
Naturally, we wonder the following question.
{\it To what extent of quantum operations does the asymptotic equation for coherence cost hold?}
We leave this problem as a future work.

\acknowledgments

We appreciate Keiji Saito, who we think almost as a co-author, for fruitful discussion and many helpful comments.
The present work was supported by JSPS Grants-in-Aid for Scientific Research No. JP19K14610 (HT), No. JP17H02861(HN) and No. JP18H03291(HN).

\clearpage

\begin{widetext}
\begin{center}
{\large \bf Supplemental Material for \protect \\ 
``Coherence-variance uncertainty relation and coherence cost for quantum measurement under conservation laws''}\\
\vspace*{0.3cm}
Hiroyasu Tajima$^{1}$ and Hiroshi Nagaoka$^{2}$ \\
\vspace*{0.1cm}
$^{1}${\small \em Yukawa institute for theoretical physics, Kyoto University
Oibuncho Kitashirakawa Sakyo-ku Kyoto city, Kyoto, 606-8502, Japan}
\\
$^{2}${\small \em University of Electro-Communications,
1-5-1 Chofugaoka, Chofu, Tokyo, 182-8585, Japan}
\end{center}

\setcounter{equation}{0}
\setcounter{lemma}{1}
\setcounter{page}{1}
\renewcommand{\theequation}{S.\arabic{equation}}

\section{Proof of the upper bound in (\ref{UandL}) in the main text}
In this section, we prove the upper bound in \eqref{UandL} in the main text.
For the convenience of the readers, below we put the inqaulity \eqref{UandL} again:
\begin{align}
\sqrt{\frac{f(0)}{2}}\frac{\|[A_S,B_S]\|}{\epsilon}-\frac{\|A_S\|}{2}\le\sqrt{I^{f}_{\epsilon,\mbox{cost}}}\le\frac{\|[A_S,B_S]\|}{2\epsilon}+\|A_S\|\label{UandL'}
\end{align}
where the upper bound holds for small $\epsilon$ satisfying $0<\epsilon\le\frac{\|[A_S,B_S]\|}{8\|A_S\|}$.

In order to show the above upper bound, we use the following lemma:
\begin{lemma}\label{bestmeasurement}
Let us take a real positive number $\xi$, and construct an implementation set $\calI=({\cal H}_E,A_E,M_E,\rho_E,U_{SE})$ as follows.
We firstly take a quantum system $S'$ whose Hilbert space has the same dimension as that of $S$.
And we also take a quantum system $E'$ whose Hilbert space is one-dimensional continuous system.
We take $E$ as the composite system $S'$ and $E'$.
We define other elements of $\calI$ as follows
\begin{align}
\rho_E&:=\psi_{\xi,E'}\otimes\ket{0}_{A_{S'}}\bra{0}_{A_{S'}}\\
U_{SE}&:=(1_S\otimes V_{S'E'})(\mathrm{SWAP}_{SS'}\otimes1_E)\\
M_{E}&:=B_{A_{S'}}\otimes1_{E'}\\
A_{E}&:=A_{S'}+A_{E'}
\end{align}
where $A_{S'}=A_S$ and $A'_E$ is the position operator of $E'$.
$\mathrm{SWAP}$ is the swap gate between $S$ and $S'$, 
$\ket{0}_{A_{S'}}$ is the ground state of $A_{S'}$, and $B_{A_{S'}}$, $\psi_{\xi,E'}$ and $V_{S'E'}$ are defined as follows:
\begin{align}
\psi_{\xi,E'}&:=C\int^{\infty}_{-\infty}e^{-\frac{x^2}{4\xi^2}}\ket{x}dx\\
V_{S'E'}&:=\sum_{jk}u_{jk}\ket{j}_{A_{S'}}\bra{k}_{A_{S'}}\otimes \gamma_{a_j-a_k}\\
\gamma_{a_j-a_k}&:=e^{-i(a_j-a_k)\hat{p}}\\
u_{jk}&:=\bra{j}_{B_{S'}}\ket{k}_{A_{S'}}\\
B_{A_{S'}}&:=\sum_{j}b_{j}\ket{j}_{A_{S'}}\bra{j}_{A_{S'}}
\end{align}
where, $\ket{j}_{A_{S'}}$ is the $j$-th eigenvector of $A_{S'}$, and $\ket{j}_{B_{S'}}$ and $b_{j}$ are the $j$-th eigenvector and eigenvalue of $B_{S'}$ that is the Hermitian operator $B_S$ on $S'$.

Then, the following inequality holds:
\begin{align}
\epsilon^2_{\calI}\le\frac{\|[B_S,A_S]\|^2}{4\xi^2}\left(1+\frac{\|A_S\|^2}{\xi^2}\right)e^{\frac{\|A_S\|^2}{2\xi^2}}
\end{align}
\end{lemma}

\begin{proofof}{Lemma \ref{bestmeasurement}}
Because of $B_{S}\otimes1_{S'}=\mathrm{SWAP}^{\dagger}_{SS'}1_S\otimes B_{S'}\mathrm{SWAP}_{SS'}$, the error $\epsilon_{\calI}(\rho_S)$ can be transformed as follows:
\begin{align}
\epsilon_{\calI}(\rho_S)^2&=\Tr[\mathrm{SWAP}_{SS'}(\rho_S\otimes\ket{0}_{A_{S'}}\bra{0}_{A_{S'}}\otimes\psi_{\xi,E'})\mathrm{SWAP}^{\dagger}_{SS'}(V^{\dagger}_{S'E'}(1_{S}\otimes B_{A_{S'}}\otimes 1_{E'})V_{S'E'}-1_S\otimes B_{S'}\otimes 1_{E'})^2]\nonumber\\
&=\Tr[(\rho_{S'}\otimes\psi_{\xi,E'})(V^{\dagger}_{S'E'}(B_{A_{S'}}\otimes 1_{E'})V_{S'E'}-B_{S'}\otimes 1_{E'})^2]\label{29}
\end{align}
Here $\rho_{S'}$ means the state $\rho_S$ on $S'$.

Let us construct tools to evaluate the righthand-side of \eqref{29}.
Due to the definition of $V_{S'E'}$, we obtain
\begin{align}
V^{\dagger}_{S'E'}(B_{A_{S'}}\otimes1_{E'})V_{S'E'}&=(\sum_{j,k}u^{*}_{jk}\ket{k}_{A_{S'}}\bra{j}_{A_{S'}}\otimes \gamma_{a_k-a_j})(\sum_{k'}b_{k'}\ket{k'}_{A_{S'}}\bra{k'}_{A_{S'}}\otimes 1_{E})\nonumber\\
&\enskip\enskip\enskip(\sum_{k'',j''}u_{k'',j''}\ket{k''}_{A_{S'}}\bra{j''}_{A_{S'}}\otimes \gamma_{a_{k''}-a_{j''}})\nonumber\\
&=\sum_{j,k,k'}b_{j}u^{*}_{jk}u_{kk'}\ket{k}_{A_{S'}}\bra{k'}_{A_{S'}}\otimes \gamma_{a_{k}-a_{k'}}\nonumber\\
&=\sum_{k,k'}B_{kk'}\ket{k}_{A_{S'}}\bra{k'}_{A_{S'}}\otimes \gamma_{a_{k}-a_{k'}}\label{31}
\end{align}
Here $B_{kk'}:=\bra{k}_{A_{S'}}B_{S'}\ket{k'}_{A_{S'}}$.
From \eqref{31} we obtain
\begin{align}
V^{\dagger}_{S'E'}(B_{A_{S'}}\otimes1_{E'})^2V_{S'E'}=\sum_{k,k',k''}B_{kk'}B_{k'k''}\ket{k}_{A_{S'}}\bra{k''}_{A_{S'}}\otimes \gamma_{a_{k}-a_{k''}}\label{32}
\end{align}
Substituting \eqref{31} and \eqref{32} for the righthand-side of \eqref{29}, we transform it into the following form.
\begin{align}
&\Tr[(\rho_{S'}\otimes\psi_{\xi,E'})(V^{\dagger}_{S'E'}(B_{A_{S'}}\otimes 1_{E'})V_{S'E'}-B_{S'}\otimes 1_{E'})^2]\nonumber\\
&=\sum_{kk'k''}B_{kk'}B_{k'k''}\rho_{k''k}(\bra{\psi_{\xi,E'}}\gamma_{a_{k}-a_{k''}}\ket{\psi_{\xi,E'}}-\bra{\psi_{\xi,E'}}\gamma_{a_{k}-a_{k'}}\ket{\psi_{\xi,E'}}-\bra{\psi_{\xi,E'}}\gamma_{a_{k'}-a_{k''}}\ket{\psi_{\xi,E'}}+1)\nonumber\\
&=\sum_{kk'k''}B_{kk'}B_{k'k''}\rho_{k''k}(e^{-\frac{(a_k-a_{k''})^2}{8\xi^2}}-e^{-\frac{(a_k-a_{k'})^2}{8\xi^2}}-e^{-\frac{(a_{k'}-a_{k''})^2}{8\xi^2}}+1)\nonumber\\
&=\sum_{kk'k''}B_{kk'}B_{k'k''}\rho_{k''k}\sum^{\infty}_{m=1}\frac{(-1)^m}{m!(8\xi^2)^{m}}((a_k-a_{k''})^{2m}-(a_k-a_{k'})^{2m}-(a_{k'}-a_{k''})^{2m})\nonumber\\
&=\sum_{kk'k''}B_{kk'}B_{k'k''}\rho_{k''k}\sum^{\infty}_{m=1}\frac{(-1)^m}{m!(8\xi^2)^{m}}(\sum^{2m}_{l=0}{}_{2m}C_{l}(-1)^{l}(a_{k}^{2m-l}a_{k''}^{l}-a^{2m-l}_ka^{l}_{k'}-a^{2m-l}_{k'}a^{l}_{k''}))\nonumber\\
&=\sum^{\infty}_{m=1}\frac{(-1)^m}{m!(8\xi^2)^{m}}\sum^{2m}_{l=0}{}_{2m}C_{l}(-1)^{l}\Tr[(A_{S'}^{2m-l}B^{2}_{S'}A_{S'}^{l}-A_{S'}^{2m-l}B_{S'}A_{S'}^{l}B_{S'}-B_{S'}A_{S'}^{2m-l}B_{S'}A_{S'}^{l})\rho_{S'}]\label{33}
\end{align}
Here we use the fact that $\bra{\psi_{\xi,E'}}\gamma_{y}\ket{\psi_{\xi,E'}}=e^{-\frac{y^2}{8\xi^2}}$ holds for arbitrary real number $y$ in the second equality.

By using $Y_{k}:=[B_{S'},A^{k}_{S'}]$, we can transform the righthand-side of \eqref{33} as follows:
\begin{align}
&\sum^{\infty}_{m=1}\frac{(-1)^m}{m!(8\xi^2)^{m}}\sum^{2m}_{l=0}{}_{2m}C_{l}(-1)^{l}\Tr[(A_{S'}^{2m-l}B^{2}_{S'}A_{S'}^{l}-A_{S'}^{2m-l}B_{S'}A_{S'}^{l}B_{S'}-B_{S'}A_{S'}^{2m-l}B_{S'}A_{S'}^{l})\rho_{S'}]\nonumber\\
&=\sum^{\infty}_{m=1}\frac{(-1)^m}{m!(8\xi^2)^{m}}\sum^{2m}_{l=0}{}_{2m}C_{l}(-1)^{l}\Tr[(A_{S'}^{2m-l}B_{S'}[B_{S'},A_{S'}^{l}]-B_{S'}A_{S'}^{2m-l}[B_{S'},A_{S'}^{l}]-B_{S'}A_{S'}^{2m}B_{S'})\rho_{S'}]\nonumber\\
&=\sum^{\infty}_{m=1}\frac{(-1)^m}{m!(8\xi^2)^{m}}\sum^{2m}_{l=0}{}_{2m}C_{l}(-1)^{l}\Tr[([A_{S'}^{2m-l},B_{S'}][B_{S'},A_{S'}^{l}]-B_{S'}A_{S'}^{2m}B_{S'})\rho_{S'}]\nonumber\\
&=\sum^{\infty}_{m=1}\frac{(-1)^m}{m!(8\xi^2)^{m}}\sum^{2m}_{l=0}{}_{2m}C_{l}(-1)^{l}\Tr[(-Y_{2m-l}Y_{l}-B_{S'}A_{S'}^{2m}B_{S'})\rho_{S'}]\nonumber\\
&=\sum^{\infty}_{m=1}\frac{(-1)^m}{m!(8\xi^2)^{m}}\sum^{2m}_{l=0}{}_{2m}C_{l}(-1)^{l}\Tr[(-Y_{2m-l}Y_{l})\rho_{S'}]\label{34}
\end{align}
Here, we use the following equality and the fact that $B_{S'}A_{S'}^{2m}B_{S'}$ is independent of $l$ in the last equality.
\begin{align}
\sum^{2m}_{l=0}{}_{2m}C_{l}(-1)^{l}=\sum^{2m}_{l=0}{}_{2m}C_{l}1^{2m-l}(-1)^{l}=(1-1)^{2m}=0
\end{align}

Therefore, 
\begin{align}
\epsilon_{\calI}(\rho_S)^2&\le\sum^{\infty}_{m=1}\frac{(-1)^m}{m!(8\xi^2)^{m}}\sum^{2m}_{l=0}{}_{2m}C_{l}(-1)^{l}\Tr[(-Y_{2m-l}Y_{l})\rho_{S'}]\nonumber\\
&=-\frac{2\Tr[\rho_{S'}[B_{S'},A_{S'}]^2]}{8\xi^2}
+\sum^{\infty}_{m=2}\frac{(-1)^m}{m!(8\xi^2)^{m}}\sum^{2m}_{l=0}{}_{2m}C_{l}(-1)^{l}\Tr[(-Y_{2m-l}Y_{l})\rho_{S'}]\nonumber\\
&\le\frac{\|[B_{S'},A_{S'}]\|^2}{4\xi^2}
+\sum^{\infty}_{m=2}\frac{1}{m!(8\xi^2)^{m}}\sum^{2m}_{l=0}{}_{2m}C_{l}\|Y_{2m-l}\|\|Y_{l}\|\label{S18}
\end{align}
Due to the definition $Y_{k}:=[B_{S'},A^{k}_{S'}]$, the following holds:
\begin{align}
Y_{k+1}=Y_{k}A_{S'}+A^{k}_{S'}[B_{S'},A_{S'}].
\end{align}
Therefore, we obtain
\begin{align}
\|Y_{2m-l}\|\|Y_l\|\le\|A_{S'}\|^{2m-2}\|[A_{S'},B_{S'}]\|^2(2m-l)l\label{S20}
\end{align}
Substituting \eqref{S20} into \eqref{S18}, we obtain the following inequalities:
\begin{align}
\epsilon_{\calI}(\rho_S)^2&\le
\frac{\|[B_{S'},A_{S'}]\|^2}{4\xi^2}
+\sum^{\infty}_{m=2}\frac{1}{m!(8\xi^2)^{m}}\sum^{2m}_{l=0}{}_{2m}C_{l}\|A_{S'}\|^{2m-2}\|[A_{S'},B_{S'}]\|^2(2m-l)l
\nonumber\\
&=\frac{\|[A_{S'},B_{S'}]\|^2}{4\xi^2}\left(1+\sum^{\infty}_{m=2}\frac{(2m-1)\|A_{S'}\|^{2(m-1)}}{(m-1)!(4\xi^2)^{m-1}}\sum^{2m-1}_{l=1}{}_{2(m-1)}C_{l-1}\right)\nonumber\\
&=\frac{\|[A_{S'},B_{S'}]\|^2}{4\xi^2}\left(1+\sum^{\infty}_{m=1}\frac{(2m+1)}{m!}\left(\frac{\|A_{S'}\|^{2}}{2\xi^2}\right)^m\right)\nonumber\\
&=\frac{\|[A_{S'},B_{S'}]\|^2}{4\xi^2}\left(1+\frac{\|A_{S'}\|^2}{\xi^2}\right)e^{\frac{\|A_{S'}\|^2}{2\xi^2}}
\end{align}
Due to $A_{S'}=A_{S}$, we obtain Lemma \ref{bestmeasurement}
\end{proofof}

\begin{proofof}{the upper bound in \eqref{UandL'}}
Let us take an arbitrary real number $\epsilon$ satisfying $0<\epsilon\le\frac{\|[A_S,B_S]\|}{8\|A_S\|}$.
To show the upper bound in \eqref{UandL'}, we only have to show that there is at least one $\calI$ satisfying 
\begin{align}
\sqrt{I^{f}_{\rho_E}(A_E)}=\frac{\|[A_S,B_S]\|}{2\epsilon}+\|A_S\|,\enskip \epsilon_{\calI}\le\epsilon.
\end{align}
Let us construct $\calI$ concretely.
We firstly define $\xi_{\epsilon}$ as follows:
\begin{align}
\xi_{\epsilon}:=\frac{\|[A_S,B_S]\|}{2\epsilon}+\|A_S\|.
\end{align}
Due to $0<\epsilon\le\frac{\|[A_S,B_S]\|}{8\|A_S\|}$, the following inequality holds
\begin{align}
\frac{\|A_S\|}{\xi_{\epsilon}}\le\frac{1}{4}.
\end{align}
Note that for $x$ satisfying $0<x\le1/4$, the following is valid
\begin{align}
(1+x)e^{x/2}\le(1+x)^2\le (1+\sqrt{x})^2.
\end{align}
Therefore, by substituting $\xi_{\epsilon}$ for $\xi$ of Lemma \ref{bestmeasurement}, we obtain $\calI^{\epsilon}$ such that
\begin{align}
\epsilon_{\calI^{\epsilon}}&\le\frac{\|[A_{S},B_{S}]\|}{2\xi_{\epsilon}}\sqrt{\left(1+\frac{\|A_{S}\|^2}{\xi^2}\right)e^{\frac{\|A_{S}\|^2}{2\xi^2}}}\nonumber\\
&\le\frac{\|[A_{S},B_{S}]\|}{2\xi_\epsilon}\left(1+\frac{\|A_{S}\|}{\xi_\epsilon}\right)\nonumber\\
&=\frac{\epsilon\|[A_{S},B_{S}]\|}{\|[A_{S},B_{S}]\|+2\epsilon\|A_S\|}(1+\frac{2\epsilon\|A_S\|}{\|[A_{S},B_{S}]\|+2\epsilon\|A_S\|})\nonumber\\
&=\frac{\epsilon\|[A_{S},B_{S}]\|(\|[A_{S},B_{S}]\|+4\epsilon\|A_S\|)}{(\|[A_{S},B_{S}]\|+2\epsilon\|A_S\|)^2}\nonumber\\
&\le\epsilon.
\end{align}
Clearly, this $\calI^{\epsilon}$ satisfies $I^{f}_{\rho_E}(A_E)=\xi^2_{\epsilon}$ and $\epsilon_{\calI^{\epsilon}}\le\epsilon$. Therefore, $\calI^\epsilon$ is the implementation that we seek.
\end{proofof}

\section{relationship between (\ref{skew-variance}) and the previous refinements of Kennard-Robertson uncertainty}
In this section, we see the relationship between \eqref{skew-variance} and the previous refinements of Kennard-Robertson uncertainty \cite{Luo-unc,Yanagi2010,Yanagi2011,Gibilisco2011,Marvian-thesis,Frowis2015}.
The previous refinements are classified in the following three types:
\begin{itemize}
\item[Type 1]{For an arbitrary $f$ satisfying $Q1$--$Q3$, the following inequality holds: 
\begin{align}
U^{f}_{\rho}(A)U^{f}_{\rho}(B)\ge f(0)^2\left|\left<[A,B]\right>\right|^2\label{type1}
\end{align}
where $U^{f}_{\rho}(A):=\sqrt{V^2_{\rho}(A)-(V_{\rho}(A)-I^{f}_{\rho}(A))^2}$.
This inequality is originally derived in \cite{Gibilisco2011} by Gibilisco and Isola.
}
\item[Type 2]{
The following inequality holds
\begin{align}
U^{f}_{\rho}(A)U^{f}_{\rho}(B)\ge f(0)\left|\left<[A,B]\right>\right|^2\label{type2}
\end{align}
for $f$ satisfying $Q1$--$Q3$ and 
\begin{align}
\frac{x+1}{2}+\tilde{f}(x)\ge 2f(x),\label{cond-Y}
\end{align}
where $\tilde{f}$ is defined as
$\tilde{f}(x):=\frac{x+1}{2}-\frac{(x-1)^2}{2}\frac{f(0)}{f(x)}.$
This inequality is originally derived in \cite{Yanagi2011} by Yanagi.
This inequality includes the original Luo's refinement \cite{Luo-unc} of the Kennard-Robertson inequality in Wigner-Yanase skew information, and its generalization for Wigner-Yanase-Dyson skew information by Yanagi \cite{Yanagi2010}.
}
\item[Type 3]{For arbitrary $f$ satisfying $Q1$--$Q3$, the following inequality holds:
\begin{align}
I^{f}_{\rho}(A)V_{\rho}(B)\ge \frac{f(0)}{2}\left|\left<[A,B]\right>\right|^2\label{type3}
\end{align}
The inequality is obtained for the case of $f=f_{WY}$ by Marvian in Ref. \cite{Marvian-thesis}, and for the case of $f=f_{SLD}$ by Fr\"{o}wis, Schmied and Gisinin in Ref. \cite{Frowis2015}.
}
\end{itemize}

Let see all of the three types obtainted from the inequality \eqref{skew-variance}.
We firstly derive \eqref{type3}.
It is easily obtaind from the fact $V_{\rho}(A)\ge V^{f}_{\rho}(A)$ for arbitrary $f$ \cite{notev}.

In order to obtain the inequalities \eqref{type1} and \eqref{type2}, we use the following relations \cite{Hansen2008,Gibilisco2011}:
\begin{align}
U^{f}_{\rho}(A)&=\sqrt{I^{f}_{\rho}(A)(V_{\rho}(A)+V^{\tilde{f}}_{\rho}(A))},\label{IJ}\\
V^{f}_{\rho}(A)&=\sum_{ij}\lambda_jf(\lambda_i/\lambda_j)|A_{ij}|^2,
\end{align}
where $A_{ij}:=\bra{i}A_0\ket{j}$, $A_0:=A-\left<A\right>_{\rho}1$, and $\{\lambda_i\}$ and $\{\ket{i}\}$ are the eigenvalues and eigenvectors of $\rho$.
From \eqref{IJ}, we obtain
\begin{align}
U^{f}_{\rho}(A)\ge \sqrt{I^{f}_{\rho}(A)V_{\rho}(A)} \ge \sqrt{I^{f}_{\rho}(A)V^{f}_{\rho}(A)} 
\end{align}
for arbitrary $f$. Therefore, we obtain \eqref{type1} from \eqref{skew-variance}.
Also, when $f$ satisfies \eqref{cond-Y}, we obtain $V_{\rho}+V^{\tilde{f}}_{\rho}(A)\ge 2V^{f}_{\rho}(A)$ as follows:
\begin{align}
2V^{f}_{\rho}(A)&=2\sum_{i,j}\lambda_j f(\lambda_i/\lambda_j)|A_{ij}|^2\nonumber\\
&\le \sum_{i,j}\lambda_{j}\left(\frac{1+\frac{\lambda_i}{\lambda_j}}{2}+\tilde{f}(\lambda_i/\lambda_j)\right)|A_{ij}|^2\nonumber\\
&=V_{\rho}+V^{\tilde{f}}_{\rho}(A).
\end{align}
Therefore, when $f$ satisfies \eqref{cond-Y},  the following inequality holds
\begin{align}
U^{f}_{\rho}(A)\ge \sqrt{2I^{f}_{\rho}(A)V^{f}_{\rho}(A)}. \label{S34}
\end{align}
The inequality \eqref{S34} and \eqref{skew-variance} directly implies \eqref{type2}.

\section{extension to framework of $G$-covariant operations}\label{sectionGcov}
In this section, we extend our framework to the resource theory of asymmetry.
As a consequence of this extension, we show that our results also clarify the amount of necessary and sufficient resource to implement non-free unitary, in the case of resource theory of $U(1)$-asymmetry.

At first, we introduce the framework of resource theory of asymmetry.
Our framework is the standard one used in Refs. \cite{Bartlett, Korezekwa-thesis, Gour, Marvian, Marvian2016, Marvian-thesis,Takagi2018, Marvian2018, Lostaglio2018,Takagibunken1}.
In the resource theory of asymmetry, the free operations are given as $G$-covariant operations that are symmetric with respect to some symmetry group $G$.
To be concrete, the $G$-covariant operation is the quantum operation $\calE$ satisfying the following equation for the unitary representation of the group $\{U_g\}_{g\in G}$:
\begin{align}
{\cal E}(U_{g}(...)U^{\dagger}_{g})=U_{g}{\cal E}(...)U^{\dagger}_{g},\enskip \forall g\in G.\label{G-cov-def}
\end{align}
Also, the free states are given as invariant states with respect to the transformation by $\{U_g\}_{g\in G}$:
\begin{align}
\rho=U_g\rho U^{\dagger}_g,\enskip \forall g\in G.
\end{align}
The above free operations (G-covariant operations) the free states (G-invariant states) satisfy the following important properties:
\begin{itemize}
\item[R1]{Every $G$-covariant operation can be realized by proper set of free state and unitary $U_g$.
In fact, for an arbitrary $G$-covariant operation ${\cal E}$ on a quantum system $A$, there is another system $B$ that has a unitary representation $\{U_g\}_{g\in G}$ of $G$ on $AB$, and we can realize ${\cal E}$ with a free state $\rho$ on $B$ and a unitary $U_{g}$ in $\{U_g\}_{g\in G}$ as follows \cite{Keyl,Marvian-thesis}:
\begin{align}
{\cal E}(...)=\Tr_{B}[U_g(...\otimes\rho)U^{\dagger}_g].
\end{align}
}
\item[R2]{We cannot transform a free state to non-free state by $G$-cavariant operation. Namely, if a state $\rho$ is $G$-invariant and a quantum operation ${\cal E}$ is $G$-covariant, then the state ${\cal E}(\rho)$ is also $G$-invariant.}
\end{itemize}

Property R2 naturally leads us to the notion that there is a kind of ``resource'' that does not increase under $G$-covariant operations.
The resource shows the degree of how far a non-free state is from free states.
There are many researches about how to measure the amount of the resource in the resource theory of asymmetry \cite{Bartlett, Gour, Marvian, Marvian2016, Marvian-thesis,Takagi2018}, and they have shown that the following properties are desirable for good measures of asymmetry. 
\begin{itemize}
\item{The measure $R$ does not increase through the $G$-covariant operations.}
\item{The measure $R$ is non-negative, and is zero if and only if $\rho$ is $G$-invariant.}
\end{itemize}
Every metric adjusted skew information used is one of well-known measures of asymmetry satisfying the above properties \cite{Marvian-thesis,Takagibunken1}.
(It also satisfies many other desirable properties including the additivity for the product states.)

Now, we have introduced the framework of resource theory of asymmetry.
It is natural thought to consider the amount of necessary and sufficient resource to implement $G$-incovariant operation by $G$-covariant operation. 
This is the key problem of resource theory of quantum channels in case of resource theory of asymmetry.
Here, we extend our framework to $G$-covariant operations, and show that our results give a complete answer to this question for the case where the implemented operation is unitary.
We also partially answer to the question for the case of non-unitary operations.

We focus on the case of $G=U(1)$.
In this case, the unitary representation $\{U_{g}\}_{g\in G}$ satisfies the following equation for some Hermitian operator $A$:
\begin{align}
U_{g}=e^{-igA}.
\end{align}
Due to this fact and Property R1, we can easily show that our framework in the main text is equivalent to the implementation of the measurement of $B_S$ with using $U(1)$-covariant operations.
In the extension, the implementation set becomes ${\cal K}:=({\cal H}_E,A_E,\rho_E,M_E,{\cal E}_{SE})$, where ${\cal E}_{SE}$ is a $U(1)$-covariant operation for $\{U_{g}\}_{g\in G}$ such that $U_g=e^{-ig(A_S+A_E)}$.
In that case, we can also define the implementation error $\epsilon_{{\cal K}}$ and coherence cost $I'^{f}_{\epsilon}$ in the same way as \eqref{maxerror} and \eqref{costformeas}.
Due to Property R1, the additivity of the metric adjusted skew informations for the product states, and the fact that the metric adjusted skew informations are zero for the G-invariant state, the inequality \eqref{lowercost} immediately gives
\begin{align}
\sqrt{I'^{f}_{\epsilon}}\ge\sqrt{\frac{f(0)}{2}}\frac{\|[A_S,B_S]\|}{\epsilon}-\frac{1}{2}\|A_S\|.
\end{align}
Also, because $U_g$ is $G$-covariant operation, the inequality \eqref{UandL} immediately gives
\begin{align}
\sqrt{I'^{f}_{\epsilon}}\le\frac{\|[A_S,B_S]\|}{2\epsilon}+\|A_S\|.
\end{align}
When $f(0)=1/2$, these upper and lower bounds give the following asymptotic equality
\begin{align}
\sqrt{I^{f}_{\epsilon}}=\frac{\|[A_S,B_S]\|}{2\epsilon}+O(\|A_S\|),\enskip \epsilon\rightarrow0.
\end{align}
Therefore, our results clarify the amount of necessary and sufficient resource to implement non-free unitary, in the case of resource theory of $U(1)$-asymmetry.

Finally, we point out that we can also show that the inequality \eqref{skew-WAY} also holds for the implementation set $\calK$, in the same way as the above discussion.

\end{widetext}

\end{document}